\documentclass[]{article}
\usepackage[a4paper, total={6in, 9in}]{geometry}
\usepackage[colorlinks,allcolors=blue]{hyperref}
\usepackage{amsmath}
\title{On the Interrelation of the Generalized Holographic Equipartition and Entropy Maximization in Kaniadakis Paradigm}
\author{
	Pranav Prasanthan$^{1}$, 
	Sarath Nelleri$^{2}$, 
	Archana K Pradeepan$^{3}$, \\
	Navaneeth Poonthottathil$^{2}$, 
	Emmanuel Tom$^{3}$ \\
	\small $^{1}$Department of Physics, Payyanur College, Payyanur-670327, India \\ 
	\small $^{2}$Department of Physics, Indian Institute of Technology, Kanpur-208016, India \\ 
	\small $^{3}$Department of Physics, Nirmalagiri College, Nirmalagiri-670701, India
}

\date{}
\begin{document}
	\maketitle
	
	\begin{abstract}
		This study examines the compatibility of the generalized holographic equipartition proposed in ref \cite{sheykhi2013friedmann} with the maximization of horizon entropy in an (n + 1)-dimensional non-flat Friedmann-Robertson-Walker (FRW) universe. Here, the entropy associated with the apparent horizon is described by Kaniadakis entropy, as well as truncated Kaniadakis entropy, which is expanded and truncated to third order when the Kaniadakis parameter $(K)$ is small, indicating minor deviations from the standard Bekenstein-Hawking entropy. Initially, we derive the conditions required for maximizing both Kaniadakis horizon entropy and truncated Kaniadakis horizon entropy. We then examine whether the generalized holographic equipartition aligns with the constraints of horizon entropy maximization. Our findings reveal that the generalized holographic equipartition is consistent with the maximization of Kaniadakis horizon entropy and truncated Kaniadakis horizon entropy in a universe with non-zero spatial curvature.
	\end{abstract}
	
	\section{Introduction}
	\label{sec:intro}
	Insights from diverse research areas, such as black hole physics \cite{bardeen1973four,bekenstein1973black,bekenstein1974generalized,hawking1975particle,hawking1976black,almheiri2013black,buoninfante2022bekenstein,hayward1998unified,bekenstein1994we}, holographic principles \cite{hooft1993dimensional,susskind1995world}, and Verlinde’s framework \cite{verlinde2011origin}, all converge to reinforce the idea that gravity and thermodynamics are fundamentally intertwined. This idea is further supported by the Jacobson's conjecture \cite{jacobson1995thermodynamics}, which claims that Einstein’s field equations can be derived by applying the Bekenstein bound and the Clausius relation together on a local Rindler causal horizon. Following this, a variety of investigations have been undertaken to explore the intricate connections between thermodynamics and gravity \cite{padmanabhan2002classical,padmanabhan2004entropy,padmanabhan2005gravity,paranjape2006thermodynamic,eling2006nonequilibrium,akbar2006friedmann,akbar2007tthermodynamic,padmanabhan2007entropy,kothawala2007einstein,padmanabhan2010thermodynamical,chakraborty2015gravitational,chakraborty2015lanczos}. These investigations were progressively expanded to include cosmic models \cite{verlinde2000holographic,cai2003holography,frolov2003inflation,cai2005first,akbar2007thermodynamic,cai2007unified,cai2007thermodynamics,sheykhi2007thermodynamical,sheykhi2007deep}, demonstrating that the Friedmann equations of the FRW universe can be framed as a manifestation of energy conservation on the apparent horizon. All of the findings mentioned highlight an interesting connection between gravity and thermodynamics. This leads us to consider gravity as the thermodynamics of spacetime. It appears that similar to thermodynamics, gravity could be an emergent phenomenon where the macroscopic geometric properties of spacetime, such as the metric and curvature, could also emerge from more fundamental entities \cite{krishna2022emergence,vt2024unified}. 
	
	While supporting the emergent gravity perspective, the studies mentioned above treat spacetime as a background geometry that already exists. The possibility that cosmic space could be emergent was first suggested by Padmanabhan \cite{padmanabhan2012emergence,padmanabhan2012emergent}. According to Padmanabhan, a pure de Sitter universe satisfies the holographic equipartition condition which relates the degrees of freedom in the bulk ($N_{bulk}$) to the degrees of freedom on the horizon ($N_{surf}$). Although our universe is not currently in a perfect de Sitter state, it is gradually transitioning towards a pure de Sitter phase that adheres to the holographic equipartition. Thus, the accelerated expansion of the universe or, similarly, the emergence of cosmic space is a result of its drive to achieve holographic equipartition \cite{krishna2017holographic}. This offers a compelling rationale for describing the spatial expansion of the universe as the emergence of cosmic space with the progression of cosmic time. Using this framework, he derived the Friedmann equation from a more fundamental law known as the law of emergence in the context of Einstein's gravity \cite{padmanabhan2012emergence}. Cai and Yang et al. extended this law of emergence to Gauss-Bonnet and Lovelock gravity theories for a spatially flat universe \cite{cai2012emergence,yang2012emergence}.  Subsequently, with minor adjustments to Cai’s approach, Sheykhi derived the Friedmann equations for universes with any spatial curvature within the frameworks of Einstein, Gauss-Bonnet, and Lovelock gravities \cite{sheykhi2013friedmann}. Further investigations in this direction
	employing Padmanabhan’s emergent space paradigm can be found in ref. \cite{padmanabhan2012physical,ling2013note,tu2013emergence,ai2013emergence,eune2013emergent,sheykhi2013emergence,ai2014generalized,ali2014emergence,chang2014friedmann,sepehri2015emergence,sepehri2016emergence,komatsu2017cosmological,sheykhi2018modified,sheykhi2021barrow,buoninfante2022bekenstein,naeem2024correction,prasanthan2024emergence,chakraborty2014evolution,dezaki2015generalized,yuan2017emergent,tu2018accelerated,mahith2018expansion,komatsu2019generalized,mathew2019emergence,hareesh2019first,padmanabhan2020gravity,mathew2022modified,krishna2022emergence,vt2022emergence,krishna2023unified,dheepika2023emergence,mathew2023emergence,vt2024unified,chen2022emergence,luciano2023emergence,chen2024emergence}.
	
	It is widely recognized that any typical macroscopic system progresses toward an equilibrium state characterized by maximum entropy \cite{callen1960thermodynamics}. Recently, Pavon and Radicella argued that the entropy of the universe, defined with the Hubble horizon as its boundary, evolves similarly to that of a typical macroscopic system, tending towards a state of maximum entropy \cite{pavon2013does}. In this regard, the authors in ref. \cite{krishna2017holographic} established that holographic equipartition corresponds to a state of maximum entropy, indicating that the law of emergence describes a universe that progresses toward a final state, the de Sitter epoch of maximum entropy. They further expanded this result to $(n+1)$ dimensional gravity theories, including Einstein, Gauss-Bonnet, and Lovelock gravities \cite{krishna2019entropy,krishna2024emergence}. Conversely, the concept of entropy maximization has been primarily examined in Einstein's gravity using the Bekenstein-Hawking entropy which arises as the black hole application of classical Boltzmann-Gibbs-Shannon (BGS) entropy in the context of Classical Boltzmann-Gibbs theory. However, its relevance is restricted when dealing with relativistic systems leading to the development of the Kaniadakis statistics or $K$-deformed statistics \cite{kaniadakis2002statistical,kaniadakis2005statistical}. The fundamental component of the Kaniadakis statistics is the Kaniadakis entropy which extends the BGS entropy by introducing a deformation parameter \( K \), known as the Kaniadakis parameter. As \(K\) approaches zero, the Kaniadakis entropy converges to the BGS entropy \cite{kaniadakis2002statistical,kaniadakis2005statistical}. This extension is motivated by the inadequacy of standard statistical models in extreme conditions, such as very high energy or density, as well as in relativistic systems \cite{kaniadakis1996generalized,kaniadakis1997non}. Here, we aim to examine the consistency of the generalized holographic equipartition \cite{sheykhi2013friedmann} with the conditions of horizon entropy maximization for an $(n+1)$-dimensional non-flat FRW universe when the entropy of the apparent horizon is described by the Kaniadakis entropy and truncated Kaniadakis entropy.
	
	It is important to highlight the recent studies examining Kaniadakis entropy and its applications within the cosmological framework scrutinized in ref. \cite{lymperis2021modified,moradpour2020generalized,hernandez2022observational,drepanou2022kaniadakis,kumar2024kaniadakis,kumar2023holographic,lambiase2023slow,sheykhi2024corrections,luciano2022modified,hernandez2022observational,hernandez2022kaniadakis,kord2023modified,salehi2023accelerating,kumar2024kaniadakis,odintsov2023holographic,nojiri2022early,nojiri2023microscopic,odintsov2024second,murali2024cosmographic,housset2024cosmological,chokyi2024cosmology}. Additionally, Black hole thermodynamics has been investigated using the Kaniadakis entropy in ref. \cite{kumar2024relativistic,abreu2021black,ambrosio2024exploring}. The recent progress and future challenges in Gravity and Cosmology using Kaniadakis statistics, showcasing its wide applicability and potential for further development in statistical modelling and theoretical physics have been demonstrated in ref. \cite{luciano2022gravity,luciano2024kaniadakis}. Motivated by the recent efforts to analyze the cosmological implications of Kaniadakis entropy, we have recently explored the relationship between the emergence of cosmic space and the first law of thermodynamics through the lens of Kaniadakis entropy \cite{prasanthan2024emergence}. We have deduced identical modified Friedmann equations in an $(n+1)$ dimensional non-flat FRW universe by employing the law of emergence \cite{sheykhi2013friedmann} and the unified first law of thermodynamics \( dE = T dS + W dV \) in conjunction with Kaniadakis entropy. This paper aims to investigate whether the law of emergence presented in ref. \cite{sheykhi2013friedmann} effectively leads to the maximization of horizon entropy in an $(n+1)$-dimensional FRW universe, specifically when the entropy related to the apparent horizon is expressed as Kaniadakis entropy and truncated Kaniadakis entropy.
	
	The remaining portion of the paper is organized as follows. The Sect. \ref{sec:2}, briefly describes the Kaniadakis entropy and truncated Kaniadakis entropy. In Sect. \ref{Sec:3}, we present the conditions for horizon entropy maximization in an ($n+1$)-dimensional non-flat FRW universe when the entropy of the apparent horizon is defined in terms of the Kaniadakis entropy and the truncated Kaniadakis entropy. Then, we check the compatibility of the generalized holographic equipartition with the conditions for horizon entropy maximization.  Finally, we conclude in  Sect. \ref{sec:4}.
	
This section explores the background and mathematical framework of the modified area entropy relation for the apparent horizon using the Kaniadakis entropy and truncated Kaniadakis entropy.
\section{Kaniadakis Entropy}\label{sec:2}
The basic component of the Kaniadakis statistics is the Kaniadakis entropy which is a one-parameter deformation of the classical BGS entropy. It arises from coherent and self-consistent relativistic statistics, while still retaining the basic
features of the classical Boltzmann-Gibbs theory \cite{kaniadakis2002statistical,kaniadakis2005statistical}. The general expression for Kaniadakis entropy is \cite{kaniadakis2002statistical,kaniadakis2005statistical}
\begin{align}\label{ks}
	{S_K = -k_B \sum_{i} n_i\ln_{\{K\}} n_i}.  
\end{align} Here $k_B$ is the Boltzmann constant, $n_i$ is the generalized Boltzmann factor for $i$-th level of a system and $\ln_{\{K\}}$ is the K-deformed logarithm, see ref. \cite{luciano2022gravity,luciano2024kaniadakis,prasanthan2024emergence} for further details about the mathematical background. Also, $K$ is the dimensionless parameter known as the Kaniadakis parameter, ranges as $-1 < K < 1$, measures deviations from the classical statistical mechanics and provides insight into the system's relativistic properties.  For applications of Kaniadakis entropy to holographic gravity \cite{hooft1993dimensional,susskind1995world,maldacena1999large}, it is useful to express Eq. \eqref{ks} in the alternative form as \cite{abreu2016jeans,abreu2017tsallis}
\begin{align}\label{777}
	S_K = -k_B\sum_{i=1}^W \frac{P_i^{1+K} - P_i^{1-K}}{2K},  
\end{align}  
where \( P_i \) represents the probability of the system being in the $i$-th state, and \( W \) denotes the total number of accessible states. Here after, we follow natural units which implies $k_B=c=\hbar=1$.

According to Einstein's gravity, the entropy of a black hole follows the Bekenstein-Hawking formula, which states that the black hole's entropy is proportional to the area of its horizon, $S_{BH}=A/4G$. Now for configurations where probabilities are equal, we assume \( P_i = {1}/{W} \), and use the fact that Boltzmann-Gibbs entropy is \( S \propto \ln(W) \), then setting \( S = S_{BH} \) leads to \( W = \exp(S_{BH}) \) \cite{moradpour2020generalized}. By substituting \( P_i = e^{-S_{BH}} \) into Eq. (\ref{777}), we obtain
\begin{align}\label{007}
	S_K=\frac{1}{K}\sinh\left(K S_{BH}\right).
\end{align}
From Eq. (\ref{007}), the relation connecting Kaniadakis entropy $(S_K)$ and apparent horizon area ($A_{n+1}$) in ($n+1$)-dimensions can be written as \cite{prasanthan2024emergence}
\begin{align}\label{aaa}
	S_K=\frac{1}{K}\sinh\left(K\frac{A_{n+1}}{4G_{eff}}\right).
\end{align}
Since the modified entropy \eqref{007} is to be close to the standard Bekenstein-Hawking value, we assume \( K \ll 1 \). Thus, expanding the Kaniadakis entropy for small \( K \) gives the truncated Kaniadakis entropy  \cite{drepanou2022kaniadakis,sheykhi2024corrections}
\begin{align}\label{bbbbb}
	S_{K} = S_{BH} + \frac{K^2}{6} S_{BH}^3 + O(K^4).  
\end{align}
In this expression, the first term is the standard Bekenstein-Hawking entropy, while the second term represents the lowest-order Kaniadakis correction. From Eq. (\ref{bbbbb}), the relation connecting truncated Kaniadakis entropy $(S_K)$ and apparent horizon area ($A_{n+1}$) in ($n+1$)-dimensions is,
\begin{align}\label{sk2}
	S_{K}=\frac{n\Omega_n \Tilde{r}_A^{n-1}}{4G_{eff}}+\frac{K^2}{6}\left(\frac{n\Omega_n \Tilde{r}_A^{n-1}}{4G_{eff}}\right)^3+ O(K^4).
\end{align}
In Eq. \eqref{aaa} and \eqref{sk2}, $A_{n+1}= n\Omega_n \tilde r_A^{n-1}$ is the area of the {apparent} horizon and ${G_{eff}}$ is the effective gravitational constant defined as ${G_{eff}=G^{\frac{n-1}{2}}}$, G being the Newtonian gravitation constant. ${G_{eff}}$ relies on the specific gravity model and reduces to conventional ${G}$ in the case of Einstein gravity \cite{tsujikawa2007matter,tian2014friedmann}. Also, in natural units ${G^{\frac{n-1}{2}}}{=}{L_P^{n-1}}$, where ${L_P}$ is the Planck length.
When \( K \to 0 \), the Eq. \eqref{aaa} and \eqref{sk2} recovers the standard Bekenstein-Hawking entropy, i.e., \( S_{K \to 0} = S_{BH} \). Also, it is important to mention that the higher order Kaniadakis correction terms in Eq. \eqref{sk2} will be neglected in the coming sections. Now, we extend these ideas to cosmology and explore the compatibility of generalized holographic equipartition \cite{sheykhi2013friedmann} with horizon entropy maximization.

\section{Generalized holographic equipartition and horizon entropy maximization}\label{Sec:3}
In ref. \cite{krishna2017holographic,krishna2019entropy,krishna2024emergence}, it has been shown that the law of emergence \cite{padmanabhan2012emergence,sheykhi2013emergence} leads to the maximization of horizon entropy for a FRW universe with any spatial curvature. Building on this, we aim to explore whether the generalized holographic equipartition proposed by Sheykhi in ref. \cite{sheykhi2013emergence} is consistent with the maximization of horizon entropy in ($n+1$)-dimensional non-flat FRW universe when the entropy associated with the apparent horizon is defined by the Kaniadakis entropy and truncated Kaniadakis entropy. 
\subsection{The law of emergence}
We begin with Padmanabahan's idea of holographic equipartition \cite{padmanabhan2012emergence}. According to Padmanabhan, the accelerated expansion of the universe or, the emergence of cosmic space is driven by the universe's tendency to achieve holographic equipartition \cite{padmanabhan2012emergence,padmanabhan2012emergent,padmanabhan2014gravity,krishna2022emergence}. The fundamental principle governing this emergence of space should connect it to the disparity between the degrees of freedom on the holographic surface and those developed in the bulk. Therefore, he proposed that the spatial expansion of the universe can be understood as the emergence of cosmic space ($dV$) over the course of cosmic time ($dt$). It can be expressed by a simple law
\begin{align}\label{p1}
	\frac{d V}{dt}= L_p^2(N_{surf}-N_{bulk}),
\end{align}
where \( N_{surf} \) represents the number of degrees of freedom on the Hubble horizon of the universe, and \( N_{bulk} \) denotes the number of degrees of freedom in the bulk. Also \( V \) is the cosmic volume and \( t \) is the cosmic time. The above law  describes a universe progressing towards a final de Sitter state, where \(N_{{bulk}}\) will ultimately equal \(N_{{surf}}\). However, using Eq. (\ref{p1}), it is not possible to derive the Friedmann equations for a non-flat (FRW) universe within alternative gravity theories \cite{cai2012emergence}. Sheykhi \cite{sheykhi2013emergence} addressed this issue by proposing that the expansion law in a non-flat universe should be generalized as follows
\begin{align}\label{shey}
	\frac{dV}{dt}=L_p^2 \frac{\tilde r_A}{H^{-1}}(N_{surf}-N_{bulk}).
\end{align}
This implies that in a non-flat universe, the volume increase continues to be directly linked to the difference between the degrees of freedom on the apparent horizon and those in the bulk. However, the proportionality factor is not just a constant, it depends on the ratio of the apparent horizon radius to the Hubble radius. The apparent horizon with radius $(\tilde{r}_A)$ is defined as \cite{hayward1998unified}
\begin{align}\label{tildere}
	\tilde r_A^2 = \frac{1}{{H^2 + \frac{k}{a^2}}}. 
\end{align}
In a spatially flat universe, where $\tilde r_A = H^{-1}$, the original relationship given in Eq. (\ref{p1}) is recovered. Therefore, the law of emergence in \( (n + 1) \)-dimensional non-flat universe can be expressed as 
\begin{align}\label{law2}
	\alpha\frac{dV}{dt}=L_p^{n-1}\frac{\tilde r_A}{H^{-1}}(N_{surf}-N_{bulk}),
\end{align}
where $V=\Omega_n \tilde r_A^n$, with $\Omega_n$ represents the Volume of unit $n-$sphere and $\alpha={n-1}/{2(n-2)}$. 

Using the definition of the Kaniadakis entropy given in Eq. \eqref{aaa}, the increase in effective volume is determined as \cite{prasanthan2024emergence},
\begin{align}\label{dv3433}
	\frac{d\tilde V}{dt}= n\Omega_n \tilde r_A^{n-1} \cosh\left(K \frac{n\Omega_n \tilde r_A^{n-1}}{4G_{eff}}\right) \dot{\tilde r}_A . 
\end{align}
With the help of the above relation, the surface degrees of freedom is proposed as \cite{prasanthan2024emergence} 
\begin{align}\label{N_surf n+11}
	N_{surf}&=\alpha \frac{n\Omega_n }{G_{eff}}\tilde r_A^{n-1}\cosh\left[K\frac{n\Omega_n \tilde r_A^{n-1}}{4G_{eff}}\right]-\alpha  \frac{ n\Omega_n }{G_{eff}}\tilde r_A^{n+1}\int_{0}^{\tilde r_A} \dfrac{\sinh\left[K\dfrac{n\Omega_n \tilde r_A^{n-1}}{4G_{eff}}\right]}{{\tilde r_A^2}} d\left[K\frac{n\Omega_n \tilde r_A^{n-1}}{4G_{eff}}\right].
\end{align}The bulk Komar energy in the case of ($n+1$)-dimensions \cite{cai2010friedmann,padmanabhan2004entropy} is given by
\begin{align}\label{N_}
	E_{Komar}=\frac{(n-2)\rho+nP}{n-2}V,
\end{align}
where the volume $V=\Omega_n \tilde r_A^n$. Then the bulk degrees of freedom becomes
\begin{align}\label{N_bulk n}
	N_{bulk}=-\frac{2E_{Komar}}{T}=-4\pi\Omega_n\tilde r_A^{n+1}\frac{(n-2)\rho+nP}{n-2},
\end{align}
where $T=\dfrac{1}{2\pi\tilde r_A}$ is the Hawking temperature. Replacing $V\to \tilde V$ and $L_p^{n-1}\to G_{eff}$, and 
substituting Eq. \eqref{dv3433}, \eqref{N_surf n+11} and \eqref{N_bulk n} in the expression \eqref{law2}, the first modified Friedmann equation with respect to the Kaniadakis entropy \eqref{aaa} in an $(n+1)$-dimensional non-flat FRW universe can be obtained as \cite{prasanthan2024emergence}
\begin{align}\label{ggggggggggggg}
	\frac{16\pi G_{eff}}{n(n-1)}\rho=\cosh\left[K\frac{n\Omega_n \tilde r_A^{n-1}}{4G_{eff}}\right]\left[H^2+\frac{k}{a^2}\right]-\int_{0}^{\tilde r_A} \dfrac{\sinh\left[K\dfrac{n\Omega_n \tilde r_A^{n-1}}{4G_{eff}}\right]}{\left(H^2+\frac{k}{a^2}\right)^{-1}} d\left[K\frac{n\Omega_n \tilde r_A^{n-1}}{4G_{eff}}\right].
\end{align}
Here we have substituted for $\tilde r_A$ from eq. \eqref{tildere}.

Now, for the definition of truncated Kaniadakis entropy \eqref{sk2}, the time progress of effective volume becomes
\begin{align}\label{dv3}
	\frac{d\tilde V}{dt}=n \Omega_n \tilde r_A^{n-1}\left(1+\alpha_{eff}\tilde r_A^{2n-2}\right)\dot{\tilde r}_A .
\end{align}
Hence, the surface degrees of freedom corresponding to the above equation can be proposed as
\begin{align}\label{N_surf n+1}
	N_{surf}=\alpha\frac{n\Omega_n }{G_{eff}}\tilde r_A^{n-1}-\alpha\frac{n\Omega_n \alpha_{eff}}{G_{eff}}\frac{\tilde r_A^{3n-3}}{n-2}.
\end{align}
The degrees of freedom present in the bulk are still described by Eq. \eqref{N_bulk n}. Substituting \( V \to \tilde{V} \) and \( L_p^{n-1} \to G_{eff} \), utilizing Eq. \eqref{dv3}, \eqref{N_surf n+1}, \eqref{N_bulk n}, \eqref{law2} and \eqref{tildere}, the first modified Friedmann equation corresponding to the truncated Kaniadakis entropy \eqref{sk2} in a non-flat (n+1)-dimensional FRW universe can be extracted as
\begin{align}\label{t22fried}
	\frac{16\pi G_{eff}}{n(n-1)}\rho= \left(H^2+\frac{k}{a^2}\right)- \frac{\alpha_{eff}}{(n-2)} \left(H^2+\frac{k}{a^2}\right)^{2-n}.
\end{align}

\subsection{Horizon entropy maximization in the Kaniadakis entropy framework}
It is a well-known principle that any ordinary, isolated macroscopic system evolves towards a state of equilibrium with the highest possible entropy, given its constraints \cite{callen1960thermodynamics}
\begin{center}
	\begin{align}\label{entr}
		\dot S \ge 0, \text{   always}
	\end{align}
\end{center}
and
\begin{center}
	\begin{align}\label{entru1}
		\ddot S < 0, \text{  in the long run}.
	\end{align}
\end{center}
In this formula, \( S \) indicates the total entropy of the universe, which can be approximated as the horizon entropy, and the dots denote derivatives concerning cosmic time \cite{krishna2024emergence}. Recently, ref. \cite{pavon2013does} demonstrated that in the context of Einstein's gravity, our universe behaves as a typical macroscopic system, progressing towards a state of maximum entropy following the above constraints. Afterwards, Krishna and Mathew broadened the approach to incorporate Gauss-Bonnet and Lovelock gravities for a universe with any spatial curvature \cite{krishna2017holographic,krishna2019entropy,krishna2024emergence}. Here, we aim to examine the maximization of horizon entropy when the entropy of the apparent horizon is represented by Kaniadakis entropy and truncated Kaniadakis entropy, within an \((n + 1)\)-dimensional FRW universe with non-zero spatial curvature. 

First, we will verify the validity of conditions \eqref{entr} and \eqref{entru1} with respect to Kaniadakis entropy \eqref{aaa}. We will now examine whether the Kaniadakis horizon entropy is maximized over time. By taking the derivative of Eq. \eqref{aaa} with respect to cosmic time, we obtain
\begin{align}\label{entropy1}
	\dot S=\frac{n(n-1)\Omega_n}{4G^{eff}}\tilde r_A^{n-2} \cosh\left(K \frac{n\Omega_n \tilde r_A^{n-1}}{4 G_{eff}}\right)\dot{\tilde r}_A.
\end{align}
To ensure that \(\dot{S} \geq 0\), \(\dot{\tilde r}_A\) in the above equation must be non-negative. Consequently, if \(\dot{\tilde r}_A \geq 0\), the entropy will not decrease.
Now, using the definition of apparent horizon radius ($\tilde r_A$) in Eq. \eqref{tildere}, the modified Friedmann equation \eqref{ggggggggggggg} corresponding to Kaniadakis entropy \eqref{aaa} can be written as 
\begin{align}\label{oky}
	\frac{16\pi G_{eff}}{n(n-1)}\rho=\cosh\left[K\frac{n\Omega_n \tilde r_A^{n-1}}{4G_{eff}}\right]\tilde r_A^{-2}-\int_{0}^{\tilde r_A} \dfrac{\sinh\left[K\dfrac{n\Omega_n \tilde r_A^{n-1}}{4G_{eff}}\right]}{\tilde r_A^{2}} d\left[K\frac{n\Omega_n \tilde r_A^{n-1}}{4G_{eff}}\right].
\end{align}
Taking the time derivative of the above equation gives
\begin{align}\label{okkkkkkk}
	\dot{\tilde r}_A =-\frac{8\pi G_{eff}}{n(n-1)} \frac{\tilde r_A^{3}}{\cosh\left[K\frac{n\Omega_n \tilde r_A^{n-1}}{4G_{eff}}\right]}\dot\rho.
\end{align}
With help of continuity equation $\dot{\rho}+nH(\rho+P)=0$ and equation of state $\omega = P/\rho$, the Eq. \eqref{okkkkkkk} can be written as,
\begin{align}\label{tilde dot 1}
	\dot{\tilde r}_A=  \frac{8 \pi G_{eff}}{n-1} \frac{\tilde r_A^3 H(1+\omega)}{\cosh\left[K\frac{n\Omega_n \tilde r_A^{n-1}}{4G_{eff}}\right]}\rho.
\end{align}
Here, $\omega$ is equation of state parameter. Recent observations suggest that our universe is evolving towards a pure de Sitter state with \(\omega \geq -1\). This ensures that \(\dot{\tilde r}_A\) remains non-negative, which in turn implies that \(\dot{S}\) is also non-negative. Next, we will examine whether this entropy reaches a maximum value in the long run, as indicated by the condition \( \ddot{S} < 0 \). To investigate this, we take the time derivative of Eq. \eqref{entropy1} which gives
\begin{align}\label{epy2}
	\nonumber\ddot S= \frac{n(n-1)\Omega_n}{4G^{eff}}\tilde r_A^{n-3}\biggl[\dot{\tilde r}_A^2\biggl(K \frac{n (n-1)\Omega_n \tilde r_A^{n-1}}{4 G_{eff}}\sinh\left(K \frac{n\Omega_n \tilde r_A^{n-1}}{4 G_{eff}}\right)+(n-2)\cosh\left(K \frac{n\Omega_n \tilde r_A^{n-1}}{4 G_{eff}}\right)\biggr)\\+\cosh\left(K \frac{n\Omega_n \tilde r_A^{n-1}}{4 G_{eff}}\right)\tilde r_A \ddot{\tilde r}_A\biggr].
\end{align}
Since $\dot{\tilde r}_A^2$ is greater than or equal to zero, \( \ddot{\tilde r}_A \) in the above equation must be negative in the asymptotic limit to satisfy the constraint \( \ddot{S} < 0 \).  Furthermore, the maximization of entropy requires that
\begin{align}\label{42}
	\nonumber\dot{\tilde r}_A^2\biggl(K \frac{n(n-1)\Omega_n \tilde r_A^{n-1}}{4 G_{eff}}\sinh\left(K \frac{n\Omega_n \tilde r_A^{n-1}}{4 G_{eff}}\right)+(n-2)\cosh\left(K \frac{n\Omega_n \tilde r_A^{n-1}}{4 G_{eff}}\right)\biggr)\\<-\cosh\left(K \frac{n\Omega_n \tilde r_A^{n-1}}{4 G_{eff}}\right)\tilde r_A \ddot{\tilde r}_A,
\end{align}
which must hold at least during the final stages of evolution. To verify this, we first compute the time derivative of Eq. \eqref{tilde dot 1}, which gives
\begin{align}\label{maxim1}
	\nonumber\ddot{\tilde r}_A =&\frac{8 \pi G_{eff}}{n-1} \frac{\tilde r_A^3 }{\cosh\left[K\frac{n\Omega_n \tilde r_A^{n-1}}{4G_{eff}}\right]}\rho\biggl[H\dot\omega-K\frac{n\Omega_n (n-1)\tilde r_A^{n-2}}{4G_{eff}} \tanh{\left[K\frac{n\Omega_n \tilde r_A^{n-1}}{4G_{eff}}\right]}\dot{\tilde r}_A H(1+\omega)+ \\&\dot H(1+\omega)+\frac{n}{2} H^2(1+\omega)^2-\frac{3n}{2}\frac{\tilde r_A^2 H^2(1+\omega)^2}{\cosh\left[K\frac{n\Omega_n \tilde r_A^{n-1}}{4G_{eff}}\right]}\int_{0}^{\tilde r_A} \dfrac{\sinh\left[K\dfrac{n\Omega_n \tilde r_A^{n-1}}{4G_{eff}}\right]}{\tilde r_A^2} d\left[K\frac{n\Omega_n \tilde r_A^{n-1}}{4G_{eff}}\right] \biggr] .
\end{align}
In the final de Sitter epoch, as \(\omega\) approaches \(-1\), all terms involving \((1 + \omega)\) vanishes. Since \(\dot{\omega}\) is always negative, this ensures that \(\ddot{\tilde r}_A < 0\) in the long term. Furthermore, in the asymptotic limit, as \(t \to \infty\) and \(\omega \to -1\), \(\dot{\tilde r}_A\) will approach zero. As a result, the inequality in \eqref{42} will be satisfied, ensuring the maximization of entropy. Therefore, within the context of Kaniadakis entropy \eqref{aaa}, the horizon entropy of a non-flat universe will not increase without bounds, thereby ensuring that entropy is maximized.

Now, we will extend this procedure to truncated Kaniadakis entropy \eqref{sk2}. To check whether truncated Kaniadakis horizon entropy is maximized, we take the time derivative of Eq. \eqref{sk2} to obtain the rate of change of truncated Kaniadakis horizon entropy as
\begin{align}\label{entroo32}
	\dot S= \frac{n(n-1)\Omega_n }{4G_{eff}} \Tilde{r}_A^{n-2}(1+ \alpha_{eff}\;\Tilde{r}_A^{2n-2}) \dot{\tilde r}_A.
\end{align}
The modified Friedmann equation \eqref{t22fried} with respect to truncated Kaniadakis entropy \eqref{sk2} can be written as
\begin{align}\label{,,}
	\frac{16\pi G_{eff}}{n(n-1)} \rho =  \tilde r_A^{-2}-\alpha_{eff}\frac{\tilde r_A^{2n-4}}{n-2}.
\end{align}
The equation above can be equivalently expressed as
\begin{align}\label{t2221f}
	\frac{16\pi G_{eff}}{n(n-1)}\rho=[1+ \alpha_{eff} \tilde r_A^{2n-2}] \tilde r_A^{-2}-\alpha_{eff} \tilde r_A^{2n-4}\frac{(n-1)}{(n-2)}.
\end{align}
Taking the time derivative of the above Eq. \eqref{t2221f} and Eq. \eqref{,,} are equivalent and will produce the same result
\begin{align}
	\dot{\tilde r}_A=-\frac{8\pi G_{eff}}{n(n-1)}\frac{\tilde r_A^3}{[1+\alpha_{eff}\tilde r_A^{2n-2}]}\dot\rho.
\end{align}     
By employing the continuity equation $\dot\rho+nH(\rho+P)=0$, we can write the above as
\begin{align}\label{tilde r 2}
	\dot{\tilde r}_A=\frac{8\pi G_{eff}}{(n-1)}\frac{\tilde r_A^3 H(1+\omega)}{[1+\alpha_{eff}\tilde r_A^{2n-2}]}\rho,
\end{align}
where $\omega$ is the equation of state parameter defined through the equation of state $\omega = P/\rho$. Since \(\omega\) is always greater than or equal to \(-1\), in an asymptotically de Sitter universe, the above equation guarantees that \(\dot{S} \geq 0\). Now to establish the condition for maximizing entropy, we compute the second derivative of the truncated Kaniadakis horizon entropy from Eq. \eqref{entroo32} as 
\begin{align}
	\ddot S= \frac{n(n-1)\Omega_n }{4G_{eff}} \tilde r_A^{n-3} \left[\dot{\tilde r}_A^2[(n-2)+ (3n-4)\alpha_{eff}\tilde r_A^{2n-2}] + (1+\alpha_{eff}\tilde r_A^{2n-2})\tilde r_A\ddot{\tilde r}_A\right].
\end{align}
To guarantee that \(\ddot{S} \leq 0\), \(\ddot{\tilde r}_A\) must be negative in the long run. Further, the constraint for maximizing horizon entropy can be immediately obtained from the above equation as
\begin{align}\label{con}
	\dot{\tilde r}_A^2[(n-2)+ (3n-4)\alpha_{eff}\tilde r_A^{2n-2}]<-(1+\alpha_{eff}\tilde r_A^{2n-2})\tilde r_A\ddot{\tilde r}_A.
\end{align}
By differentiating Eq. \eqref{tilde r 2}, we obtain
\begin{align}\label{maxim2}
	\nonumber\ddot{\tilde r}_A= \frac{8\pi G_{eff}}{n-1}\frac{\tilde r_A^3}{[1+\alpha_{eff}\tilde r_A^{2n-2}]}\rho\biggl[\frac{n}{2}{H^2(1+\omega)^2}- \frac{3n}{2}\frac{\tilde r_A^2 H^2(1+\omega)^2}{[1+\alpha_{eff}\tilde r_A^{2n-2}]}\alpha_{eff}\tilde r_A^{2n-4}\frac{n-1}{n-2}\\-\frac{\alpha_{eff}(2n-2)\tilde r_A^{2n-3}}{[1+\alpha_{eff}\tilde r_A^{2n-2}]} H(1+\omega)\dot{\tilde r}_A +\dot H (1+\omega)+H\dot\omega\biggr].  
\end{align}
Here, we substituted for $\rho$ from Eq. \eqref{t2221f}.
As \(\omega\) approaches \(-1\) in the final stage, all terms involving \((1 + \omega)\) disappear. Consequently, \(\ddot{\tilde r}_A\) will be negative in the final stage because \(\omega\) is a decreasing function. Additionally, since \(\dot{\tilde r}_A^2 \to 0\) as \(t \to \infty\), the constraint in Eq. \eqref{con} will be satisfied during the final stage. Hence, within the framework of truncated Kaniadakis entropy \eqref{sk2}, the horizon entropy of a non-flat universe is maximized in the final stage.

Therefore, from Eq. \eqref{maxim1} and \eqref{maxim2}, we can conclude that horizon entropy is maximized in the final stage when the associated entropy of the apparent horizon is defined by the Kaniadakis entropy and the truncated Kaniadakis entropy.

\subsection{The law of emergence and maximization of horizon entropy}
Here, we will check the compatibility of the generalized holographic equipartition proposed in ref. \cite{sheykhi2013emergence} with the conditions for maximization of both Kaniadakis horizon entropy and truncated Kaniadakis horizon entropy.

To begin, we will investigate whether the law of emergence leads to the maximization of horizon entropy when the entropy of the apparent horizon is expressed using the Kaniadakis entropy \eqref{aaa}. By combining Eq. \eqref{dv3433} and \eqref{entropy1}, we can relate the rate of emergence to the rate of change of entropy as
\begin{align}
	\frac{d\tilde V}{dt}=\frac{4G_{eff} \tilde r_A}{n-1}\dot S.
\end{align}
Now the law of emergence in Eq. \eqref{law2} could be expressed as
\begin{align}\label{looo}
	\dot S = \frac{(n-2)H}{2}(N_{surf}-N_{bulk}). 
\end{align}
To ensure that \(\dot{S} \geq 0\), the holographic discrepancy \(\left(N_{{surf}} - N_{{bulk}}\right)\) in the equation above must be non-negative. We will now examine whether the definitions of \(N_{{surf}}\) and \(N_{{bulk}}\) guarantee that \(\dot{S} \geq 0\). Recalling, \eqref{N_surf n+11} we have,
\begin{align}\label{N_surf n+111}
	\nonumber N_{surf}&= \frac{n(n-1)}{2(n-2)} \frac{\Omega_n \tilde r_A^{n-1}}{G_{eff}}\cosh\left[K\frac{n\Omega_n \tilde r_A^{n-1}}{4G_{eff}}\right]\\&- \frac{n(n-1)}{2(n-2)}  \frac{ \Omega_n \tilde r_A^{n+1}}{G_{eff}}\int_{0}^{\tilde r_A} \dfrac{\sinh\left[K\dfrac{n\Omega_n \tilde r_A^{n-1}}{4G_{eff}}\right]}{{\tilde r_A^2}} d\left[K\frac{n\Omega_n \tilde r_A^{n-1}}{4G_{eff}}\right].
\end{align}
Using the modified Friedmann equation \eqref{okkkkkkk} corresponding to Kaniadakis entropy \eqref{aaa}, the bulk degrees of freedom \eqref{N_bulk n} can be written as
\begin{align}\label{koio}
	\nonumber N_{bulk}=-\frac{n(n-1)}{2(n-2)G_{eff}H}\Omega_n \tilde r_A^{n-2}\dot{\tilde r}_A \cosh\left[K\frac{n\Omega_n \tilde r_A^{n-1}}{4G_{eff}}\right]+ \frac{n(n-1)}{2(n-2)} \frac{\Omega_n \tilde r_A^{n-1}}{G_{eff}}\cosh\left[K\frac{n\Omega_n \tilde r_A^{n-1}}{4G_{eff}}\right]\\- \frac{n(n-1)}{2(n-2)}  \frac{ \Omega_n \tilde r_A^{n+1}}{G_{eff}}\int_{0}^{\tilde r_A} \dfrac{\sinh\left[K\dfrac{n\Omega_n \tilde r_A^{n-1}}{4G_{eff}}\right]}{{\tilde r_A^2}} d\left[K\frac{n\Omega_n \tilde r_A^{n-1}}{4G_{eff}}\right].
\end{align}
From the above Eq. \eqref{N_surf n+111} and \eqref{koio}, the holographic discrepancy can be expressed as
\begin{align}\label{difference}
	N_{surf}-N_{bulk}= \frac{n(n-1)}{2(n-2)G_{eff}H}\Omega_n \tilde r_A^{n-2}\cosh\left[K\frac{n\Omega_n \tilde r_A^{n-1}}{4G_{eff}}\right]\dot{\tilde r}_A .
\end{align}
To satisfy the condition \( N_{{surf}} - N_{{bulk}} \geq 0 \), \( \dot{\tilde r}_A \) in the expression above must always be non-negative. If we assume the universe is asymptotically de Sitter, \( \dot{\tilde r}_A \) will be greater than or equal to zero, thereby ensuring the non-negativity of \( \dot{S} \). Next, we will examine whether the horizon entropy reaches its maximum value over time. To do this, we calculate the second derivative of horizon entropy using Eq.\eqref{looo} as
\begin{align}\label{bnmn}
	\ddot S= \frac{(n-2)\dot H}{2}(N_{surf}-N_{bulk})+\frac{(n-2)H}{2}\frac{d}{dt}(N_{surf}-N_{bulk}).
\end{align}
In the final de Sitter stage, \( N_{{bulk}} \) approaches \( N_{{surf}} \), causing the first term in the expression above to vanish. Since \( N_{{bulk}} \) does not exceed \( N_{{surf}} \), \( N_{{surf}} - N_{{bulk}} \) will always be positive and approach zero in the final stage. Consequently, we have
\begin{align}
	\frac{d}{dt}(N_{surf}-N_{bulk})<0,
\end{align}
which ensures $\ddot S <0$. In essence, the universe is trying to reduce the holographic discrepancy, thereby ensuring that \( \ddot{S} \) remains non-positive over time. Additionally, combining Eq. \eqref{difference}, and \eqref{bnmn}, the condition for \( \ddot{S} \) to be non-positive can be directly derived as
\begin{align}\label{oka}
	\nonumber\dot{\tilde r}_A^2\biggl(K \frac{n(n-1)\Omega_n \tilde r_A^{n-1}}{4 G_{eff}}\sinh\left(K \frac{n\Omega_n \tilde r_A^{n-1}}{4 G_{eff}}\right)+(n-2)\cosh\left(K \frac{n\Omega_n \tilde r_A^{n-1}}{4 G_{eff}}\right)\biggr)\\<-\cosh\left(K \frac{n\Omega_n \tilde r_A^{n-1}}{4 G_{eff}}\right)\tilde r_A \ddot{\tilde r}_A.
\end{align}
This is equivalent to the constraint given in Eq. \eqref{42}, which we derived earlier for maximizing the Kaniadakis horizon entropy. Therefore, the law of emergence ensures the maximization of horizon entropy when the entropy associated with the apparent horizon is given by the Kaniadakis entropy \eqref{aaa}.

We will now examine whether the law of emergence results in the maximization of horizon entropy when the apparent horizon's entropy is defined using the truncated Kaniadakis entropy \eqref{sk2}. By combining equations \eqref{dv3} and \eqref{entroo32}, we can relate the rate of emergence to the rate of change of entropy as
\begin{align}
	\frac{d\tilde V}{dt}=\frac{4G_{eff} \tilde r_A}{n-1}\dot S.
\end{align}
Then we can express the law of emergence in Eq. \eqref{law2} as
\begin{align}\label{oiiiiip}
	\dot S = \frac{(n-2)H}{2}(N_{surf}-N_{bulk}).
\end{align}
The holographic discrepancy \( N_{{surf}} - N_{{bulk}} \) in the above equation can be determined using the relations defined in Eq. \eqref{N_surf n+1} and \eqref{N_bulk n}. Upon substitution, we obtain 
\begin{align}\label{diff22}
	N_{surf}-N_{bulk}=\frac{n(n-1)}{2(n-2) G_{eff}H}\Omega_n \tilde r_A^{n-2}[1+\alpha_{eff}\tilde r_A^{2n-2}]\dot{\tilde r}_A.
\end{align}
For \( \dot{S} \) to be non-negative, the condition \( \dot{\tilde r}_A \geq 0 \) must be satisfied, as indicated by the above equations. This requirement is generally met in an asymptotically de Sitter universe. Next, we will determine the second derivative of horizon entropy by differentiating Eq. \eqref{oiiiiip}, which gives
\begin{align}\label{jjjjjj}
	\ddot S= \frac{(n-2)\dot H}{2}(N_{surf}-N_{bulk})+\frac{(n-2)H}{2}\frac{d}{dt}(N_{surf}-N_{bulk}).
\end{align}
At the final stage of evolution, \( N_{{bulk}} \) will equal \( N_{{surf}} \), causing the first term in the equation above to vanish. Since the universe is attempting to minimize the holographic discrepancy, we have the condition
\begin{align}
	\frac{d}{dt}(N_{surf}-N_{bulk})<0.
\end{align}
This implies that \( \ddot{S} \) will be non-positive in the long run. Also, by substituting Eq. \eqref{diff22} in Eq. \eqref{jjjjjj}, we obtain the constraint for the non-positivity of \( \ddot{S} \) in the long term as 
\begin{align}
	\dot{\tilde r}_A^2[(n-2)+ (3n-4)\alpha_{eff}\tilde r_A^{2n-2}<-(1+\alpha_{eff}\tilde r_A^{2n-2})\tilde r_A\ddot{\tilde r}_A.
\end{align}
This is identical to the inequality in equation \eqref{con}, which we derived earlier as the constraint for maximizing truncated Kaniadakis horizon entropy. Therefore, the law of emergence leads to the maximization of horizon entropy in a non-flat universe when the entropy of the apparent horizon is defined by the truncated Kaniadakis entropy \eqref{sk2}. 

Based on the above discussion, we can conclude that the generalized holographic equipartition is consistent with the maximization of both Kaniadakis horizon entropy and truncated Kaniadakis horizon entropy for an ($n+1$)-dimensional non-flat FRW universe.

\section{Conclusions}\label{sec:4}
In this study, we examined the compatibility of the generalized holographic equipartition with the conditions for maximization of Kaniadakis horizon entropy and truncated Kaniadakis entropy. Initially, we determined the constraints resulting from maximizing Kaniadakis horizon entropy and truncated Kaniadakis entropy and then examined whether a non-flat universe is consistent with those constraints. Using the continuity equation, it is demonstrated that an asymptotically de Sitter universe with \( \omega \geq -1 \) evolves toward a state of maximum horizon entropy in the Kaniadakis entropy framework. Then, we examined whether the law of emergence
results in the maximization of horizon entropy in a ($n+1$)-dimensional non-flat FRW universe.
According to the law of emergence, the condition \( \dot{S} \geq 0 \) leads to \( (N_{{surf}} - N_{{bulk}}) \geq 0 \). Since \( N_{{bulk}} \) does not exceed \( N_{{surf}} \), \( \dot{S} \) will always be non-negative. Conversely, applying the constraint \( \ddot{S} < 0 \) to the law of emergence results in the condition \( \frac{d}{dt} (N_{{surf}} - N_{{bulk}}) < 0 \). As the universe aims to reduce the holographic discrepancy \( N_{{surf}} - N_{{bulk}} \), this inequality will hold in the long run. Furthermore, we have demonstrated that these conditions are consistent with the horizon entropy maximization in the Kaniadakis framework. Our findings demonstrate that both the law of emergence and the horizon entropy maximization impose the same final constraints. In this sense, the emergence of cosmic space can be viewed as a tendency to maximize horizon entropy in a non-flat universe when the entropy associated with the apparent horizon is in the form of Kaniadakis entropy or truncated Kaniadakis entropy. 

In this study, we limited our analysis to equilibrium thermodynamics. However, the literature includes numerous studies that address non-equilibrium thermodynamics, especially in the context of non-standard entropies. Therefore, it is important to look into the implications of the Kaniadakis entropy framework within a non-equilibrium setting which will be addressed in future work.

	\section*{Acknowledgements}
	One of the authors Sarath Nelleri acknowledges the Indian Institute of Technology Kanpur for
	providing the Institute postdoctoral fellowship. The authors acknowledge the AI assisted copy editing to improve the quality of the manuscript. The authors thank Dr. Sreejith E.K. for fruitful discussions on the manuscript.


\end{document}